\documentclass[prd,aps,showpacs,showkeys]{revtex4-1}
\usepackage{bm,amsfonts,latexsym,amsmath,amssymb,amsbsy,graphicx}

\newcommand{\bb}{\begin{eqnarray}}
\newcommand{\ee}{\end{eqnarray}}
\newcommand{\ba}{\begin{align}}
\newcommand{\ea}{\end{align}}

\begin{document}

\title{\bf  On elastic scattering amplitude of  planar charged fermions  in a constant  magnetic field}
\author{V.R. Khalilov}\email{khalilov@phys.msu.ru}
\affiliation{Faculty of Physics, M.V. Lomonosov Moscow State University, 119991,
Moscow, Russia}

\begin{abstract}
The elastic scattering amplitude (ESA) is obtained in the one-loop approximation of the  2+1 dimensional quantum electrodynamics (QED$_{2+1}$) for  planar charged fermions in an external constant magnetic field. We obtain the elastic scattering amplitude in the corresponding massive theory and then discuss and calculate the massless limit of ESA.  The imaginary part of ESA are related to the total probability of photons emission by charged fermions in the considered magnetic field. Simple analytical formulas are   obtained when planar   charged fermions in initial states occupy  high excited  Landau levels.
\end{abstract}



\maketitle

\section{Introduction}

A number of condensed-matter quantum effects point to the existence of quantum-mechanical  systems that can be adequately described by the two-dimensional Dirac equation.
Planar charged fermions governed by Dirac equation  with external
electromagnetic fields attract considerable interest in connection with problems of the quantum Hall effect \cite{2}, high-temperature superconductivity \cite{3} as
well as graphene (see, e.g., \cite{6} and \cite{7,8,9}).
Electron states near the Fermi surface in graphene were usually described by the two-dimensional massless Dirac equation  \cite{7,10,11,12} which allows to consider graphene as the condensed matter analog of the quantum electrodynamics in 2+1 dimensions \cite{ggvo,datdc}.
This spatially two-dimensional quantum system in the so-called supercritical Coulomb potential becomes unstable which manifests  in the appearance of fermion quasi-stationary states  with  negative ``complex" energies \cite{6,7,8,9,10,11,vp11a,as11b,as112,15}.

In the quantum electrodynamics these fermion quasi-stationary states  are directly associated with the fermions pair creation from vacuum and the imaginary part of corresponding scattering amplitude determines the creation probability of charged fermion pair by supercritical Coulomb potential.
Fundamental phenomenon such as the electron-hole pair production due to the instability of the quantum electrodynamics vacuum in the supercritical Coulomb potential is presently within experimental reach in a graphene with charged impurities (see, \cite{gr0,gr1,gr2}).
The creation probability of planar charged fermions pair in the supercritical Coulomb field were obtained in \cite{khmpl} (see, also, \cite{khep}).

The various quantum electrodynamics effects with planar charged fermions in the presence of  external electromagnetic fields can be of considerable interest.
Significant role plays the fermion mass operator matrix elements of which determine the elastic scattering amplitude of the fermion interacting with the quantum electrodynamics vacuum in the background field, i.e., the radiative correction to the fermion energy.
The one-loop radiative correction to an electron energy in the ground state in a constant  magnetic field in the 2+1 dimensional quantum electrodynamics  was calculated in \cite{TBZh} and the one-loop  electron self-energy in the topologically massive  QED$_{2+1}$ at finite temperature and density was obtained in \cite{ZhEm}.

The radiative correction to the  lowest Landau level of massive  Dirac electron
in an external magnetic field in a thin medium simulating graphene  was investigated in the one-loop approximation of the so-called reduced QED$_{3+1}$ on a 2-brane in \cite{machet}.
It should be emphasized that in the model of the reduced QED$_{3+1}$ on a 2-brane, fermions are confined to a plane, but  the electromagnetic interaction between them is spatially three-dimensional \cite{4,4a}.  In \cite{machet} it was shown for the first time that the above radiative correction    does not vanish at the limit $m\to 0$.
The radiative corrections to the lowest Landau level of (massive and massless) charged planar fermion  for the cases of the pure QED$_{2+1}$ and the reduced QED$_{3+1}$ on a 2-brane were studied in \cite{khgen}. The radiative corrections are real in these models and are related to the radiative  shift of fermion energy in the ground state in the magnetic field.

The elastic scattering amplitude of charged fermions in excited states
in external electromagnetic fields is of value because the imaginary part of ESA is related to the total probability of photon emission by charged fermion in external fields.
In massless quantum electrodynamics the radiation problem has peculiar features, occurring when photon is emitted in the direction of the charge momentum \cite{8w}.
In the framework of classical electrodynamics the problem of electromagnetic radiation from  massless charges was studied in  \cite{1a,2a, 3l,4k} (see, also, \cite{jko}).
The problem of radiation in massless scalar quantum
electrodynamics in an external magnetic field was solved in \cite{gal}.
D.V. Gal'tsov \cite{gal} showed that synchrotron radiation from massless charges
in the external constant magnetic field is quantum effect,
investigated all main features of radiation for high Landau excitation levels and
 predicted the effect of magnetic generation of the square of mass
in the linear order in $e^2$. In work \cite{asit} it was predicted  a new type of the neutrino electromagnetic radiation due to the hypothesized neutrino electric millicharge
\cite{abps}.

In present work we study the elastic scattering amplitude in the one-loop approximation  of QED$_{2+1}$ for  charged planar fermions occupying  excited Landau levels in an external constant magnetic field.  The imaginary part of ESA of a planar charged massless fermion in excited state in the magnetic field  is nonzero. Simple expressions for the elastic scattering amplitude are obtained for  high excited Landau levels.

We shall adopt the units where $c=\hbar=1$.

\section{$E_p$ representation}

In 2+1 dimensions, as is well known (see, e.g., \cite{fwil,rpsg,scse}), the algebra of Dirac's $\gamma^{\mu}$ matrices can be represented by the two-dimensional Pauli matrices $\sigma_j$. Therefore, it is normally considered that there are
two types of fermions in 2+1 dimensions that correspond to the signature of
 these  Dirac matrices
\bb
\eta=\frac{i}{2}{\rm Tr}(\gamma^0\gamma^1\gamma^2)=\pm 1.
\label{sgn}
\ee
The two signs of $\eta$ correspond to the two nonequivalent representations of the Dirac matrices
\bb
 \gamma^0= \eta\sigma_3,\quad \gamma^1=i\sigma_1,\quad \gamma^2=i\sigma_2, \eta=\pm 1.
\label{1spin}
\ee
The Dirac equation for a fermion of  mass $m$ and charge $e=-e_0<0$ in an external electromagnetic field in 2+1 dimensions is written in the covariant form as
\bb
 (\gamma_{\mu}P^{\mu}-m)\Psi(z)=0,\label{dieq},
\ee
where
$P^{\mu} = p^{\mu} - eA^{\mu}\equiv (P_0, P_1, P_2)$ is the generalized fermion momentum operator (a three-vector),  $p^{\mu}= (i\partial_t,-i\partial_x,-i\partial_y)\equiv (p_0, p_1, p_2)$ and $z^{\mu}=(z^0,z^1,z^2)\equiv (t,x,y)$. We take the vector potential in the Cartesian coordinates in the Landau gauge   $A_0=0,\quad A_1=0,\quad A_2=Bx$, then the magnetic field is defined as $B=\partial_1A_2-\partial_2A_1\equiv F_{21}$, where $F_{\mu\nu}$ is the electromagnetic field tensor.

The squared Dirac operator is
\bb
(\gamma P)^2= P_0^2-P_1^2-P_2^2 -e\sigma_3 B.
\label{dirop}
\ee
The matrix function $E_p$, introduced by Ritus \cite{virit} for the case of 3+1 dimensions,  satisfies the equation
\bb
(\gamma P)^2E_p= p^2E_p,
\label{epf}
\ee
where the eigenvalue $p^2$ can be any real number and in the magnetic field considered $E_p$ also is an eigenfunction of the operators
\bb
i\partial_t E_p=p_0 E_p,\quad -i\partial_y E_p=p_2 E_p,\quad (P_1^2+P_2^2-e\sigma_3  B)E_p=2|eB|k E_p, \quad k=0, 1 \ldots \quad .
\label{epf1}
\ee
It is obvious that $p_0, p_2$ and $2|eB|k$ label the solutions of Dirac equation (\ref{dieq}).  As operators (\ref{epf}) and (\ref{epf1}) commute with $\sigma_3$, their eigenfunctions $E_p$  also differs by the eigenvalues    $\zeta=\pm 1$ of $\sigma_3$ and they  can be labelled by the index
 $n=k-{\rm sign}(e_0B)\zeta/2 -1/2,\quad n= 0,1, \ldots $.
The functions $E_p(z)$ are orthogonal
\bb
\int \bar E_{p'}(z)E_{p}(z)d^3z= (2\pi)^3\delta(p'_0-p_0)\delta(p'_2-p_2)\delta_{k'k},
\label{epze}
\ee
and satisfy the completeness condition
\bb
\sum_{k=0}^{\infty}\int \frac{dp_0 dp_2}{8\pi^3} E_{p}(z) \bar E_{p}(z')=\delta(z'-z),
\label{epcomp}
\ee
where $\bar E_p=\gamma^0 E_p^{\dagger}\gamma^0$ and $E_p^{\dagger}$ is the Hermitian conjugate matrix function.
It can be verified the relation
\bb
(\gamma P)E_p=E_p(\gamma \bar p),
\label{imprel}
\ee
where  the three-vector $\bar p_{\mu}=(\bar p_0,\bar p_1,\bar p_2)=(p_0, 0,{\rm sign}(e_0B)\sqrt{2|eB|n}$ depends only on the "dynamic" quantum numbers $p_0$ and $n$ and $p^2=\bar p^2$.
Relation (\ref{imprel}) implies that the solution of Dirac equation (\ref{dieq}) can be represented in the form
\bb
 \Psi(z)=E_{p}(z)u_{\bar p},\label{soldieq},
\ee
where $u_{\bar p}$ is a spinor satisfying the "free" Dirac equation
$(\gamma \bar p-m)u_{\bar p}=0$, $E_p$ is taken at $p^2=m^2$
and
\bb
p_0\equiv E_n=\sqrt{m^2+2n|eB|}
\label{Ll}
\ee
is the energy eigenvalues (the Landau levels).  It should be also emphasized that
$\Psi(z)$ are not eigenfunctions of $\sigma_3$.

Quantum states of a 2+1 massive Dirac fermion also may be characterized by the spin number   (see, for example, \cite{4c,vrks} and \cite{ggt}). For this in \cite{vrks}
it was suggested to consider only one type of fermions and to interpret the number $\eta = ±1$  as a quantum number that explicitly describes the
spin of a fermion in 2+1 dimensions, which is a pseudoscalar with respect to
Lorentz transformations.
The conserved spin operator for the Dirac equation in an external homogenous electromagnetic field in 2+1 dimensions is defined similarly to the 3+1 case (about the 3+1 case see \cite{tkp}) as  the even part of the corresponding
spin operator. The even part of the operator $S = \sigma_3/2$ is singled out using the sign operator ${\cal H}/|{\cal H}|$  and  the eigenvalues of the conserved spin operator thus obtained is  $s=\eta m/(2|E|)$, where $E$ is the eigenvalue of Hamilton operator ${\cal H}$.

It is helpful to write the spinor $u_{\bar p,\eta}$ that satisfies the "free" Dirac equation with $\gamma^0=\eta \sigma_3$
\bb
u_{\bar p,\eta}=\left(
\begin{array}{c}
u_1\\
u_2
\end{array}\right)
=\frac{1}{\sqrt{2E_n}}\left(
\begin{array}{c}
\sqrt{E_n+\eta m}\\
{\rm sign}(e_0B)\sqrt{E_n-\eta m}
\end{array}\right), \label{doubm}
\ee
A massless fermion  does not have a spin degree of freedom in 2+1 dimensions \cite{jacna} in accordance with the eigenvalues $s$ at $m=0$ and Eq. (\ref{doubm}).

\section{Mass operator and elastic scattering amplitude of charged fermion
in an external constant magnetic field}

The renormalized one-loop mass operator of a charged massive fermion in an external constant magnetic field in 2+1 dimensions in the $E_p$ representation
\bb
M(p,p',B)=\int d^3z \int d^3z' \bar E_p(z)M(z,z',B)E_p'(z')
\label{mop1}
\ee
was obtained  in \cite{khgen}.
In (\ref{mop1})
\bb
M(z,z', B)=ie^2\gamma^{\mu}S^c(x,x',B)\gamma^{\nu}S_{\mu\nu}(x-x') - ie^2\gamma^{\mu}S_{\mu\nu}(x-x') {\rm tr}(\gamma^{\nu} S^c(x,x',B))
\label{MO}
\ee
is the one-loop mass operator in the coordinate representation, $S^c(x,x',B)$ is the causal fermion Green's function in the external field and
$S_{\mu\nu}(x-x')$ is the photon propagation function.
We need the mass operator calculating with using the "effective" internal photon propagator for the reduced QED$_{3+1}$ on a 2-brane. Then,
the photon  is allowed to also propagate in the "bulk" \cite{4}. Needed mass operator is diagonal $M(p,p',B)=\delta^3(p-p')M(p,B)$ and can be represented in the form
\bb
M_c(p,B) =-\frac{e^2}{8\pi^{1/2}}\int_0^{\infty}ds \int_0^1 \frac{du}{\sqrt{1-u}}\left[\exp(-i\Phi)\frac{|eB|}{\Delta\sin(|eB|su)}\left(
(3MI -P_0\sigma_3)e^{-i\phi}-i\bar p_2 \sigma_2 s(1-u)\right)-\right.\nonumber\\
\left.-\frac{3mI-\sigma_3p_0(1-u)-i\sigma_2\sqrt{p_0^2-m^2}s(1-u)}{s}|_{\gamma p=m}\exp(-im^2su^2)
 \right],\phantom{mmmmmmmmmmm}
\label{mopn}
\ee
where
$$
\Phi=m^2su-p_0^2su(1-u)+[(p_0^2-m^2)/|eB|]\phi,
$$

$$
\Delta=\sqrt{(|eB|s(1-u))^2 +(1+|eB|s(1-u)\cot(|eB|su))^2},
$$

$$
 \phi=\arctan\frac{|eB|s(1-u)}{[1+|eB|s(1-u)\cot(eBsu)]},
$$
$$
M=ip_0(1-u)\sin(|eB|su) + m\cos(|eB|su),\quad P_0=p_0(1-u)\cos(|eB|su)+im\sin(|eB|su)
$$
and $I$ is the unit two-column matrix.
 Equation (\ref{mopn}) is valid for the on shell mass operator
(i.e. for  $E_p$ taken at $p^2=m^2$) for all Landau levels $n$.

In the QED$_{2+1}$ unlike the QED$_{3+1}$ the mass operator of charged massive fermion in an external constant magnetic field can contain one more term due to the nonzero vacuum fermion current
$ie_0{\rm tr}(\gamma^{\nu} S^c(x,x',B))$ induced by the background magnetic field for each type of fermions (see \cite{27}). It reads
\bb
M_v(p, B)=\frac{3e^2}{8\pi}\frac{eB}{m}\eta\sigma_3.
\label{vcm}
\ee
Such a term does not appear  if we interpret $\eta=\pm 1$ as a quantum number describing the fermion spin in an external electromagnetic field and on this account we must sum in $\eta$ in the second term of Eq.(\ref{MO}).
The elastic scattering amplitude  of a charged planar massive fermion in the order in $e^2$ is determined with the matrix element of mass operator (\ref{mop1}) as
\bb
A_{p',p}=- u_{\bar p',\eta'}^{\dagger} M(p,p',B)u_{\bar p,\eta}.
\label{amp1}
\ee

The purpose of the paper is to show that the imaginary part of ESA  for a charged planar massless fermion in an external constant magnetic field is nonzero. So, bearing this in mind let us find   ESA for the massive case in the limit $p_0\gg m$. Conserving only   terms  $\sim 1, m/p_0$ in the factor outside the exponent in the integrand, we obtain
\bb
A_{p,p}(n,B,\eta,\eta') =\frac{e^2}{8\pi^{1/2}}\int_0^{\infty}ds \int_0^1 \frac{du}{\sqrt{1-u}}\left[\exp(-i\Phi)\frac{|eB|}{\Delta\sin(|eB|su)}\left(
(3M -P_0m(\eta+\eta')/2p_0)e^{-i\phi}+\phantom{mmmmmmmmmmm}\right.\right.\nonumber\\
\left.\left. +\bar p_2s(1-u)m(\eta-\eta')/2p_0  \right)-
\frac{3m-m^2(1-u)(\eta+\eta')/2p_0+ m(\eta-\eta')s(1-u)/2}{s}|_{\gamma p=m}\exp(-im^2su^2)
 \right].\phantom{mmmmmmmmmmm}
\label{ampm}
\ee
We see that ESA in the massive case depends linearly on the spin numbers in the initial and final spin states.  Spin terms in Eq. (\ref{ampm}) vanish after  summing over two final spin states and averaging over the initial spin state. All these spin terms vanish in the massless case.

With neglect of terms $\sim m/p_0$ in Eq. (\ref{ampm}), we obtain
\bb
A_{p,p}(n,B) =\frac{3e^2}{8\pi^{1/2}}\int_0^{\infty}ds \int_0^1 \frac{du}{\sqrt{1-u}}\left[ \frac{|eB|}{\sin(|eB|su)}\frac{\exp(-i\Phi_0)}{\Delta}M_0e^{-i\phi}\right].
\label{easm0}
\ee
Here $\Phi_0=-2n|eB|su(1-u)+2n\phi$, $M_0=i\sqrt{2|eB|n}(1-u)\sin(|eB|su)$.
This expression does not contain the fermion mass and can directly be obtained  from Eq.(\ref{ampm}) at $m=0$. It is the elastic scattering amplitude of a charged planar massless fermion in the external field.
 In accordance with
the so-called  optic theorem \cite{blp} the imaginary part of ESA is related to the total probability of photons emission
\bb
2{\rm Im}A_{p,p}=W(E,B).
\label{prob1}
\ee

We will be interested in the case of high initial Landau levels $n\gg 1$.
For this case  the integrals over $s$ can be evaluated by the method of stationary phase (see, for instance, \cite{olw}). Let us introduce a variable $x=|eB|su$ and then expand
a phase $\Phi_0$, $\Delta$ and $\phi$ in powers of $x$ so that
\bb
\Phi_0\approx 2n P(u)x^3, P(u)=(1-u)^2u/3, \quad \Delta\approx 1/u, \quad \phi\approx x(1-u)
\label{expan}
\ee
and
\bb
A_{p,p}(n,B) =\frac{3ie^2}{8\pi^{1/2}}\sqrt{2|eB|n}\int_0^{\infty}dx \int_0^1 du{\sqrt{1-u}}e^{-i2nP(u)x^3}e^{-ix(1-u)}.
\label{easm1}
\ee
For $n\gg 1$ the integrals over $x$ are estimated  by the method of stationary phase. In particular, the integral $I(n)=\int_0^{\infty}dx \exp[-i2nP(u)x^3]$ in the leading approximation in $1/n^{1/3}$ can be evaluated as
\bb
I(n) \sim e^{-i\pi/6}\Gamma(1/3)3^{-1}[2nP(u)]^{-1/3},
\label{est}
\ee
where $\Gamma(z)$ is the Euler gamma function.
The integral over $u$ is reduced to the Euler beta function $\beta(5/6,2/3)$ (see, for instance, \cite{GR})
\bb
\int_0^1 du (1-u)^{-1/6}u^{-1/3}=\beta(5/6,2/3).
\label{intu}
\ee

Thus, we finally obtain
\bb
{\rm Im}A_{p,p}(n,B)= \frac{e^2}{4}3^{1/3}\Gamma(5/6)(|eB|E)^{1/3}=W(E,B)/2,
\label{imp}
\ee
where $E=\sqrt{2n|eB|}$ is the energy eigenvalue of massless charged fermion.
The real part of ESA describes the radiative correction to the energy eigenvalue $E$.
In the leading approximation in $1/n^{1/3}$ this correction is
\bb
{\rm Re}A_{p,p}(n,B)= \frac{e^2}{4}3^{-1/6}\Gamma(5/6)(|eB|E)^{1/3}.
\label{reap}
\ee
It should be noted that the radiative correction to the Landau level $n\gg 1$ (\ref{reap}) substantially differs  from that to the Landau level $n=0$ (see, \cite{khgen}). One can easily understand that Eqs. (\ref{imp}) and (\ref{reap}) give main mass-independent  terms in ESA in the massive case in the so-called ultra-quantum limit $|eB|E/m^3\gg 1$.

\section{Resume}

In this work we calculate the elastic scattering amplitude of planar  charged  fermions in an  external constant  magnetic field in the one-loop  approximation of the  2+1 dimensional quantum electrodynamics. We show that the imaginary part of ESA of a planar charged massless fermion in the considered magnetic field is nonzero for initial excited Landau levels $n$. Simple analytical expressions for the elastic scattering amplitude are obtained for  high excited Landau levels $n\gg 1$. As regards low initial Landau levels $n\neq 0$  the calculations are not so simple and can be performed numerically.
Here, it will be noted that photons emission of charged massive particles (electrons) in low initial Landau levels in an external magnetic field was analyzed in \cite{vab,ofd,vgb} (see, also, references therein).

Results obtained here seem to be likely to be related to graphene in the presence of external constant magnetic field (normal to the monolayer graphene).

The author is grateful to I. V. Mamsurov for useful conversations.

\end{document}